\documentclass[lnicst]{svmultln}
\usepackage{graphicx}
\usepackage{csquotes}
\usepackage[utf8x]{inputenc}
\usepackage{makeidx}  
\begin{document}
\mainmatter              
\title{Sense-Deliberate-Act Cognitive Agents for Sense-Compute-Control Applications in the Internet of Things \& Services}
\titlerunning{BDI Agents for the IoTS}  
%
\author{Amir H. Moin}
\authorrunning{Amir H. Moin}   
%
%
\institute{fortiss, An-Institut Technische Universität München, Munich, Germany\\
\email{moin@fortiss.org}}

\maketitle              

\begin{abstract}        
In this paper, we advocate Agent-Oriented Software Engineering (AOSE) through employing Belief-Desire-Intention (BDI) intelligent agents for developing Sense-Compute-Control (SCC) applications in the Internet of Things and Services (IoTS). We argue that not only the agent paradigm, in general, but also cognitive BDI agents with sense-deliberate-act cycle, in particular, fit very well to the nature of SCC applications in the IoTS. However, considering the highly constrained heterogeneous devices that are prevalent in the IoTS, existing BDI agent frameworks, even those especially created for Wireless Sensor Networks (WSNs), do not work. We elaborate on the challenges and propose possible approaches to address them.
\end{abstract}
\section{Introduction}\label{introduction}
The Internet of Things and Services (IoTS) is an expanded version of the Internet containing not only the components of the Internet such as Web 2.0, but also constrained embedded devices such as sensors and actuators. As a crucial infrastructure for Cyber Physical Systems (CPS), in which the physical world merges with the virtual world of cyberspace \cite{Broy2012a}, the IoTS is believed to have sufficient power to trigger the next (i.e., fourth) industrial revolution. However, it turns out that with this great power an enormous degree of complexity in software systems is inevitable. This is due to a number of reasons such as the tremendously large number of \enquote{things} (i.e., devices), heterogeneity of things and services as well as the variety of communication protocols.

Sense-Compute-Control (SCC) applications \cite{Patel+2014} are a typical group of applications in the IoTS. A SCC application senses the environment (e.g., temperature, humidity, light, UV radiation, etc.) through sensors, performs some computation (often decentralized, i.e., distributed) and finally prompts to take one or more actions through actuators (very often sort of control) in the environment. There exist two main differences between these applications in the IoTS and the similar ones in the field of Wireless Sensor and Actuator Networks (WSAN), a predecessor of the field of the IoTS. First, the scale of the network is quite different. While WSANs typically have several hundreds or thousands of nodes, SCC applications in the IoTS may have several millions or billions of nodes. Second, the majority of nodes in a WSAN are more or less similar to each other. However, here in the IoTS we have a wide spectrum of heterogeneous devices, ranging from tiny sensor motes with critical computational, memory and energy consumption constraints to highly capable servers for cloud computing. Heterogeneity is a property inherited from another predecessor field, known as Pervasive (Ubiquitous) computing. \cite{Patel+2011}

One of the recent paradigms in software engineering that helps in dealing with complexity is Agent-Oriented Software Engineering (AOSE). In fact, it is the intersection of the field of Multi-Agent Systems (MAS), the successor of Distributed Artificial Intelligence (DAI), with software engineering. Although the research field of MAS is only about two decades old, many methodologies have been proposed in this area. They could be mainly categorized into two groups \cite{Sturm+2004, Weiss2001}: software engineering (e.g., Gaia, Prometheus and O-MaSE) and knowledge engineering (e.g., CoMoMAS, MAS-CommonKADS and Tropos). Our focus here is on the former category, a field that is known as Agent-Oriented Software Engineering (AOSE). 

AOSE helps in dealing with complexity in software development through raising the level of abstraction via the higher level concepts of agents, roles, organizations, collaboration, interaction protocols, etc. Yet, cognitive agents (e.g., BDI agents) raise the level of abstraction beyond that level. Moreover, MASes are intrinsically peer-to-peer distributed systems with asynchronous communication often based on message-passing with unicast, multicast and broadcast possibilities. This is a model which fits perfectly fine to the nature of SCC systems in the IoTS. Because, unlike web applications in the Internet, where the client/server model is prevalent, in the IoTS the peer-to-peer network topology (either pure or hybrid) makes more sense. Also, the asynchronous communication based on message-passing is much more scalable and efficient comparing other alternatives for distributed systems such as synchronous Remote Procedure Calls (RPC), in which the client should block (i.e., busy-wait) until the server finishes the processing of its request.

There exist several research works that have proposed BDI MAS approaches for the IoTS, WSANs, etc. However, considering the fact that the majority of heterogeneous things in the IoTS are highly constrained devices in terms of computational power, memory space, energy consumption, etc., to our knowledge, none of the proposed BDI-based approaches could work in practice for the SCC applications in the IoTS. Therefore, as the main contribution, we propose possible approaches for addressing this problem.

The paper is structured as follows. In Section \ref{bdi-scc}, we explain the idea of employing BDI agents for SCC applications in the IoTS, elaborate on the practical technical challenges and propose two possible approaches for addressing those challenges. This is followed by a brief review of the related work in this area in Section \ref{related-work}. Finally, we conclude and mention our future work in Section \ref{conclusion}.

\section{BDI Agents for SCC Applications}\label{bdi-scc}
Intelligent agents have a number of key properties which distinguish them from other software paradigms. Some of them are listed below \cite{Sturm+2004}.
\begin{enumerate}
\item \textit{Autonomy:} An agent can perform its task without any supervision.
\item \textit{Reactiveness:} An agent reacts in a timely manner to the changes in its environment.
\item \textit{Proactiveness:} An agent is not only reactive, but also proactive in the sense that it is goal-oriented and pursues its own goals.
\item \textit{Sociality:} An agent interacts with other agents often through message-passing. A group of agents collaborating with each other in order to achieve a common goal form a society known as organization.
\end{enumerate}

In a broad sense, the AOSE research area can be divided into two groups \cite{Braubach+2005}. The first group is more concentrated on the MAS paradigm as middleware (e.g., Java Agent DEvelopment Framework (JADE) \cite{Bellifemine+2003}), while the second one is about reasoning-oriented agent frameworks (e.g., the Jadex BDI system \cite{Braubach+2005}). Our discussion here is more related to the latter category, i.e., reasoning.

As mentioned in the previous section, cognitive agents provide a very high level of abstraction for hiding complexity in AOSE. A number of cognitive architectures for agents have been proposed in the literature. One of the well-known software models for cognitive intelligent agents is Belief-Desire-Intention (BDI). BDI is not only simple to implement, but also corresponds nicely to the way that people talk about the human behavior in psychology \cite{Braubach+2005}.

\begin{figure}[h]
    \centering
    \includegraphics[width=0.6\textwidth]{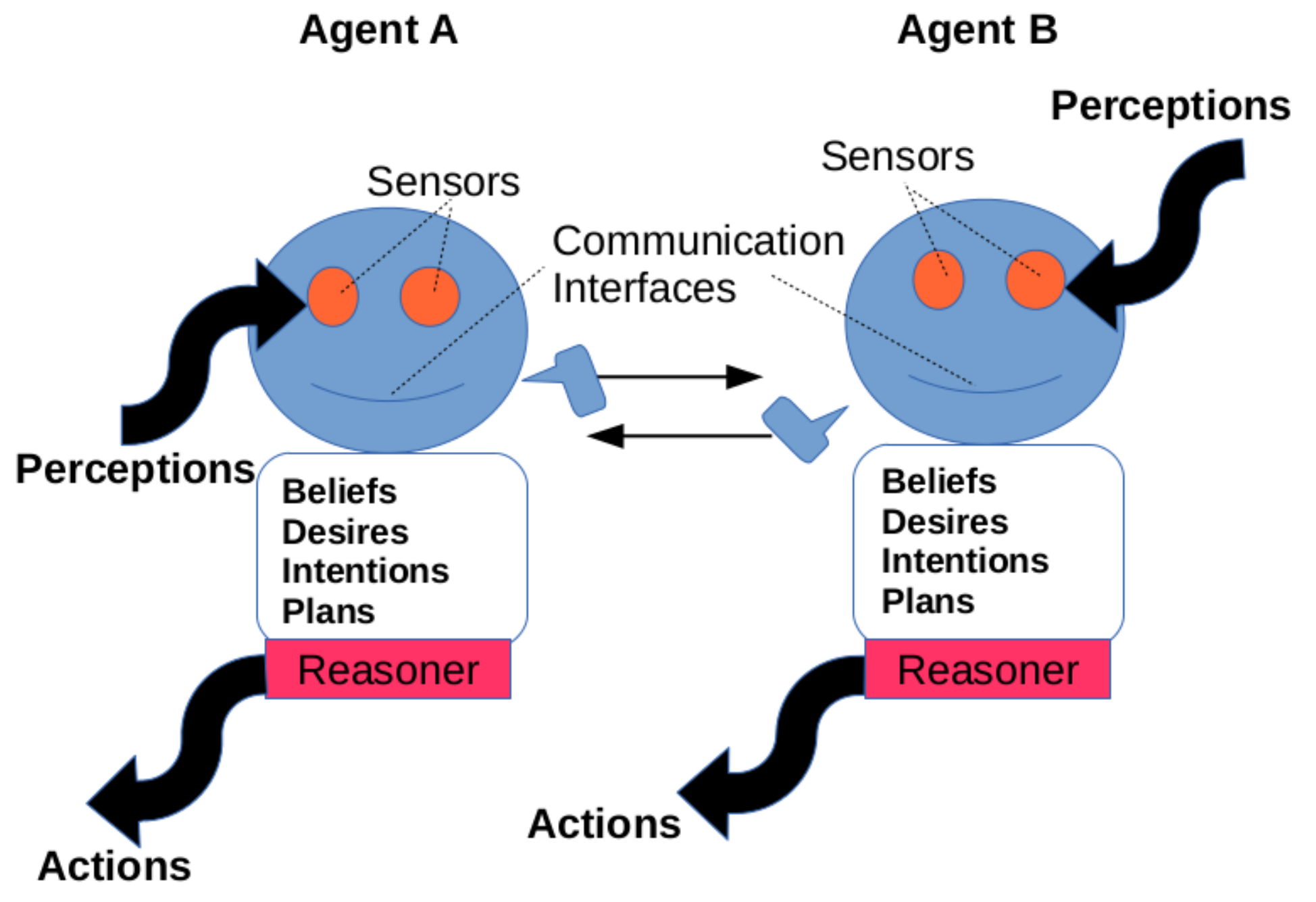}
    \caption{Belief-Desire-Intention (BDI) agents, figure inspired by \cite{HuhnsSingh1998}}
    \label{fig:bdi}
\end{figure}

As depicted in Figure \ref{fig:bdi}, a BDI agent has a set of beliefs, a set of desires, a set of intentions and a set of plans. Beliefs are informational attitudes about the surrounding environment (i.e., the local personal view of the agent from the world that it perceives through its sensors) as well as the internal state of the agent itself. Desires are motivational attitudes that represent the wishes of the agent. These wishes may even be conflicting with each other. A consistent subset of desires is called goals. Goals can have different types such as achieve-goals (for achieving a state) or maintain-goals (for maintaining a state over a period of time). Another important component, is the set of plans. Each plan consists of a sequence of actions for achieving a goal and can even contain subgoals. Hence, other plans are needed to achieve the subgoals of one plan. This way, a hierarchy of plans is formed. Moreover, intentions are those goals that the agent is currently committed to achieve them based on its plans.\cite{Braubach+2005}

The BDI model is implemented in a number of MAS frameworks, namely PRS, dMARS, JACK Intelligent Agents\texttrademark, JAM, AgentSpeak, 3APL, Dribble, Coo-BDI, Brahms, and Jadex. So far, BDI agents have proven to be very useful in various areas such as industrial automation, simulation, air traffic control systems, e-commerce systems, virtual environments, online multi-player games, etc. Among the customers of only one of the BDI frameworks, called dMARS (a predecessor of Jadex), are NASA (space shuttle malfunction handling), AirServices, Thomson Airsys (air traffic control), Daimler Chrysler (supply chain management, resource and logistics management) and Hazelwood Power (process control).\cite{Mascardi+2005}

From the above explanation, it is clear that the BDI model fits very well to the logic of SCC applications in the IoTS. Moreover, as mentioned in Section \ref{introduction}, MASes are intrinsically very close to the nature of the IoTS. Therefore, for our application domain, i.e., SCC applications in the IoTS, we advocate the BDI model for MASes. State-of-the-art BDI systems such as the free open source Jadex BDI agents framework \cite{Braubach+2005} are much more mature and advanced comparing the initial BDI frameworks. Moreover, Jadex, which supports both stationary and mobile agents, can be deployed on different middleware such as JADE. JADE is a free open source FIPA\footnote{http://www.fipa.org/}-compliant middleware for the development and runtime execution of peer-to-peer applications based on the MAS paradigm \cite{Bellifemine+2003}. Furthermore, there exist an extension of JADE known as JADE-LEAP, for mobile platforms (e.g., Android).

However, given the prevalence of highly resource-constrained things in the IoTS, such as tiny sensor motes with only few kilobytes of memory and extremely limited computational (i.e., CPU) as well as power (i.e., electrical energy) resources, neither Jadex nor any other state-of-the-art BDI agent framework, that we are aware of, could work for such limited platforms in the IoTS. In order to address this problem, we propose three possible approaches. First, one may tailor JADE to work atop very compact and efficient virtual machines of Java-like byte-code languages that are especially designed for motes such as the IBM Mote Runner. Second, an existing C/C++ BDI framework such as the classic BDI reasoner called PRS (e.g., OpenPRS) may be evolved in order to work for this purpose. Finally, the more feasible solution is to develop a BDI framework from scratch, especially for SCC applications in the IoTS that works for constrained devices too, similar to the non-BDI Mobile-C, a mobile agent platform for mobile C/C++ agents on constrained embedded devices, or Agila \cite{Fok+2005} and actorNet \cite{Kwon+2006} which both run on TinyOS for Wireless Sensor Network (WSN) applications.

Last but not least, one drawback of the BDI model is that its reasoning power often comes at the cost of increased computational overhead. However, by using the BDI model more judiciously and making correct design choices, one can overcome this problem. For a nice practical example on this issue for using BDI with Brahms language based on efficient design choices in contrast to inefficient ones at NASA Ames Research Center, please refer to \cite{Wolfe+2008}.

\section{Related Work}\label{related-work}
MASes are widely used in the field of Wireless Sensor and Actuator Networks (WSAN). ActorNet \cite{Kwon+2006} is an agent-based platform with an extremely lightweight interpreter that can operate on low-power sensor nodes with as little as 4KB RAM. The programming language for agents is functional and very similar to the Scheme programming language in terms of its syntax. The communication model is based on message passing. ActorNet is specifically designed for mobile agents on Mica2 sensor motes running TinyOS. Moreover, Agila \cite{Fok+2005} is an agent-based middleware for mobile agents on Mica2, MicaZ, and Tmote Sky motes running TinyOS. It has an assembly-like programming language with proprietary ISA for agents \cite{Aiello+2010}.

In contrast, MAPS (Mobile Agent Platform for Sun SPOT) \cite{Aiello+2011} and AFME (Agent Factory Micro Edition) \cite{Muldoon+2006} are Java-based frameworks. The agent model in the former is Finite-State-Machines (FSM), while the latter is based on the BDI model. However, they can only run on Sun SPOT (Sun Small Programmable Object Technology), but not on the more constrained tiny sensor motes that are much more prevalent in the IoTS.

\section{Conclusion \& Future Work}\label{conclusion}
Although there exist overlapping areas in the two young research fields of the IoTS and Multi-Agent Systems, where the latter is only about one decade older than the former, the research and development communities are still isolated from each other. We believe that the R\&D work in the field of the IoTS could benefit a lot from the advances in MASes. Similarly, the IoTS could become an interesting testbed for MASes.

In this paper, we advocated Agent Oriented Software Engineering (AOSE), in particular Belief-Desire-Intention (BDI) cognitive intelligent agents, for Sense-Compute-Control (SCC) applications in the Internet of Things and Services (IoTS). Considering the specific requirements and the constrained devices which are prevalent in the IoTS, we elaborated on challenges for pursuing this aim and proposed possible approaches for addressing them. The implementation and validation of the proposed approaches remained as future work.
\bibliographystyle{IEEEtran}
\bibliography{IEEEabrv,myrefs}
\end{document}